\documentclass[aps,pre,showpacs,preprintnumbers,amsmath,amssymb,floatfix]{revtex4}
\usepackage{amsmath,graphicx,amsfonts}
\usepackage{dcolumn}

\newcommand{\be}{\begin{equation}}
\newcommand{\ph}{\vec{\phi}} 
\newcommand{\ee}{\end{equation}}

\begin{document}
{
\title{\bf Scattering of topological solitons on holes and barriers}

\author{Bernard Piette and W.J. Zakrzewski\footnote{also at
 Max Planck Institute for the Physics
of Complex Systems, N{\"o}thnitzer Stra{\ss}e 38, 01187 Dresden, Germany}} 
\affiliation{Department of Mathematical Sciences, University of Durham,
 Science Laboratories, South Road, Durham DH1 3LE, England}

\author{Joachim Brand}
\affiliation{Max Planck Institute for the Physics
of Complex Systems, N{\"o}thnitzer Stra{\ss}e 38, 01187 Dresden, Germany}

\date{\today}

\begin{abstract}

We study the scattering properties of topological solitons on
obstructions in the form of holes and barriers.
We use the 'new baby Skyrme' model in (2+1) dimensions and we model
the obstructions by making the coefficient of the baby skyrme model
potential - position dependent. We find that that the barrier leads to the
repulsion of the solitons (for low velocities) or their complete transmission
(at higher velocities) with the process being essentially elastic.
The hole case is different; for small velocities the solitons are trapped
while at higher velocities they are transmitted with a loss of energy.
We present some comments explaining the observed behaviour.
\end{abstract}


\maketitle


\section{Introduction}
Consider a moving particle encountering a potential barrier or a
potential hole.  Then, the barrier slows down the particle and if the
energy of the particle is too low relative to the barrier's height,
the particle is reflected.  In the potential hole case the particle
speeds up in the hole and is always transmitted.

However, consider now the same process in quantum mechanics; then in
both cases we have reflection and transmission, the relative
magnitudes of which depend on the parameters of the potential and on
the energy of the particle.

Recently, some work has been done on the scattering properties of
solitons of the nonlinear Schr\"odinger model
\cite{Cao,Goodman,Sakaguchi,Flach}. It was pointed out in
Ref.~\cite{Joachim} that the scattering of solitons at low
velocities resembles a classical particle in the sense that the
soliton maintains its integrity and follows a well-defined trajectory,
with the difference that, nevertheless, the soliton can be reflected by
a potential hole. Such reflection is not possible in classical
mechanics.


Motivated by these results,
we have decided to look at this problem in the case of topological
solitons asking ourselves the question as to
whether this case will resemble more the classical particle 
case or the quantum mechanical systems?

To look at this more carefully we have chosen to study this problem on
the example of a (2+1) dimensional system so that we could see the
effects of the transverse direction and to allow for the radiation
waves, if they are generated, to escape more easily. The model we have
chosen is the baby Skyrme model discussed in detail in many
publications \cite{baby1,baby2}. The mutual interactions of
Skyrmions in scattering events have been a topic of long standing
interest and are summarized in
Refs.~\cite{scat1,scat2,scat3}. Experimental realizations of
topological solitons are possible in spinor Bose-Einstein condensates
\cite{bec1,bec2}.
  
The Lagrangian density of the ``baby Skyrme'' model contains three terms, from left to right:
the pure ${\cal S}\sp2$ sigma model, the Skyrme and the potential terms:

\begin{equation}\label{lagrangian} 
{\cal L}=\partial_{\mu}\ph\cdot\partial^{\mu}\ph
-\theta_{S}\left[(\partial_{\mu}\vec{\phi} \cdot
\partial^{\mu}\vec{\phi})^{2} -(\partial_{\mu}\vec{\phi} \cdot
\partial_{\nu}\vec{\phi}) (\partial^{\mu}\vec{\phi} \cdot
\partial^{\nu}\vec{\phi})\right] -  V(\ph)
\end{equation}
where
\be
V(\phi) = \mu(1-\phi_3^2).
\ee 
The vector $\ph$ lies on the unit sphere ${\cal S}^{2}$
hence $\ph\cdot\ph=1$.\\
To have a finite potential energy the field at spatial infinity 
is required to  go to $\phi_3=\pm 1$, $\phi_1=\phi_2=0$.
In this work we choose ``the vacuum'' to be defined 
as $\phi_3=+1$. 

Note that our boundary condition has defined a one-point compactification of $R_2$,
allowing us to consider $\ph$ on the extended plane 
$R_2\bigcup{\infty}$  topologically equivalent to  ${\cal S}^{2}$.
In consequence, the field configurations are maps   
\begin{equation} 
\vec{\phi} \, : {\cal S}\sp2 \longrightarrow {\cal S}\sp2.  
\end{equation}
which can be labelled by an integer valued topological index $Q$:
\begin{equation} 
Q=\frac{1}{4\pi}\int  \,\vec{\phi}\cdot
\left(\partial_{x}\vec{\phi}\times\partial_{y}\vec{\phi}\right)\,dx\,dy.
\end{equation}
As a result of this non-trivial mapping the model has 
topologically nontrivial solutions 
which describe ``extended structures", which have 
been called baby skyrmions.
A soliton is then the simplest field configuration $\ph$
corresponding to $Q=1$ which minimises
the total energy and  which can be calculated from $\cal L$ (\ref{lagrangian}) 
by taking all terms in it with positive signs. 
This soliton field configuration has to be found numerically;
however, as discussed in previous papers  \cite{baby1,baby2},
this problem reduces to having to solve an ordinary differential
equation for a ``profile function'' $f(r)$, where $r$
is the radial distance from the position  of the soliton.

Note that the soliton is exponentially localised and its asymptotic
behaviour is controlled by $\mu$.

In the next section we describe a possible way of introducing
an ``obstruction'' into our model and discuss the results of our studies. In Sec. III 
we present the results of our
 numerical simulations.   Sec. IV presents some concluding remarks.           

\section{Potential Obstruction}

There are various ways of introducing a potential hole or a potential
barrier. However, given that the soliton field, strictly speaking, is
never zero, even though it vanishes exponentially as we move away from
its position, this potential has to be introduced in such a way that
it does not change the ``tail'' of the soliton {\it i.e.} it has to
vanish when $\phi_3=1$.  A possible way to do this is to add an extra
term to the Lagrangian which vanishes when $\phi_3=1$. Of course,
there are many possible choices of such terms but given that our
Lagrangian already contains a term with such a property we exploit
this fact and choose to add $\alpha(1-\phi_3^2)$ in some region of $x$
and $y$. We choose this term to be independent of $y$ so that the
obstruction on the potential energy landscape, located in some finite
region of $x$, say at positive $x$, resembles a trough in the ``hole''
case or a dam in the ``barrier'' case. Then sending the soliton from
a point well away from this obstruction, {\it i.e.} initially placed
at some sufficiently negative $x$, in the positive $x$ direction, we
can study the effects of the obstruction.

In our numerical simulations we have chosen the obstruction to be constant
in a small range of $x$; this effectively corresponds to taking
$\mu$ in the original Lagrangian to be given by $\mu_0$
for $x$ in the range of the obstruction and $\mu_1$ elsewhere.

The case when $\mu_0> \mu_1$ corresponds to a barrier (dam), 
and when $\mu_0< \mu_1$ we have a hole (trough).
We have performed many numerical simulations of such systems, varying
both the sign and value of $\mu_0-\mu_1$ and the velocity of the incoming soliton.
We have also checked that, initially, the solitons were far enough
from the obstruction so that the incoming solitons can be considered to be free.

In the next section we present some results of our simulations.
Both the cases of the hole and of the barrier have been 
studied and, as we shall see, they have produced very different results.
The scattering by the barrier was found to be  elastic, with hardly
any radiation being generated in the process; by comparison the motion
through the trough was inelastic with an interesting pattern of the 
decrease of velocity of the transmitted soliton.

\section{Numerical Simulations}

\subsection{General comments}
We have performed most of our simulations on a 350$\times350$ square grid with lattice
spacing being 0.15.  Thus the lattice extended from -26.5 to +26.5 in each direction.
The soliton was initially placed at $x=-12$, $y=0$. Its size was determined by the 
choice of parameters, $\theta_s$ and $\mu$ which were chosen as
$\theta_s=0.5$ and $\mu_1=0.6$. This produced a soliton which was essentially
localised to  about 40 lattice points 
in each direction {\it i.e.} to a region of $6\times 6$ in real space.
 Thus, placing the soliton at $x=-12$ there was no
problem with any boundary effects.

Next we put our obstruction from $x=0$ to $x=30*0.15=4.5$. We considered
various values of the height of the obstruction. As they all led to
qualitatively similar results we performed most of our simulations for
$\mu_1-\mu_0=0.5$ and $-0.1$.
The same is true when we varied the width
of the barrier although changing the width changed the value of $v_{cr}$.

All our simulations were performed using a 4th order Runge-Kutta method
with the field $\ph$ being rescaled every few iterations. The time step of our
simulations was taken to be 0.001. 
We used fixed boundary conditions and later we used also the absorbing boundary 
conditions at the edges of the lattice. This we generated by successively
decreasing the magnitude of ${d\ph\over dt}$ at the last 5 rows and columns of the lattice.

To generate the time dependence we calculated the initial configurations 
with the soliton located at $x=-12$ and $x=-11.9$ and we defined 
${d\ph\over dt}$ as being proportional to the difference between these two $\ph$'
orthogonalised with respect to $\ph$ at $x=-12$. Varying the constant of 
proportionality we changed the initial velocity. This is 
an approximate 
way of introducing the initial condition corresponding to a moving soliton,
which is a very good approximation for small values of velocities, when 
the relativistic effects are negligible. Using this method we have to calibrate the 
velocity - to check directly what the initial velocity is.
Another way to proceed would involve introducing the correct time dependence
into the initial ansatz (with Lorentz factors etc) and then calculating 
${d\ph\over dt}$ directly from this expression. Given the fact that the initial
configurations are calculated numerically and involve some extrapolations 
we chose the easier option mentioned above. Our results show that our procedure
was exceptionally good; the moving soliton varied its height very little and it did not
generate any perceptible amount of radiation.
 
\subsection{Barrier}

First we have considered the effects of a barrier. Hence, between
$x=0$ and $x=4.5$,  we put $\mu_1-\mu_0=-0.1$. Then we placed the initial
soliton at various values of $x$ and calculated its energy without altering
the shape of the soliton. This has told us what the barrier is like  as seen by
the soliton. 
The calculated energy plot is shown in Fig.~1 
\begin{figure}
\begin{center}
\includegraphics[width=8cm]{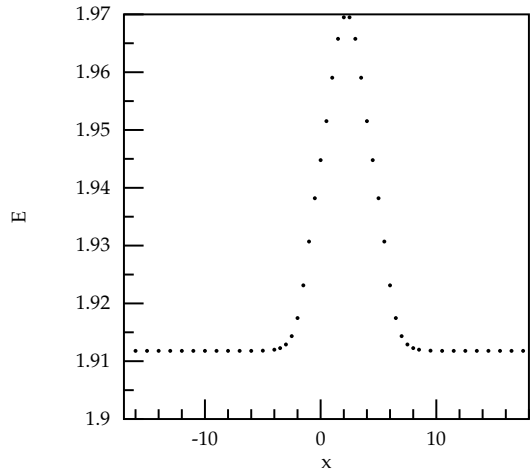}
\caption{  Energies of our basic soliton as a function of its 
position}
\end{center}
\end{figure}

We see that, although the original barrier is in the shape of a square
barrier, the soliton perceives the barrier as a smooth hump or hill.
This is, of course, due to the finite size of the soliton. As seen in Fig.~1, the barrier 
effects stretch from $x\sim -4$ to $x\sim 10$.

The soliton placed at $x=-12$ should be far away from the effects of
the barrier. Next we performed a series of simulations placing the
soliton at $x=-12$ and sending it towards the barrier at various
speeds.  We have found extremely elastic behaviour in all cases. For
low velocities ($v<v_{cr}\sim 0.2427$) the soliton bounced off the
barrier while for $v>v_{cr}$ it was transmitted. At velocities close
to $v_{cr}$ the soliton slowly climbed the ascending slope of the
effective barrier of Fig.~1, either getting over it ($v>v_{cr}$) or
falling back.  In all the cases the process was elastic.  This was
seen through the plots of energy density, where we did not see any
significant radiation energy. Also the values of the final velocity of
the soliton were essentially the same as the initial ones, indicating
that no significant energy loss had taken place.  To calculate the
initial velocity we used the values of the time when the soliton
reached $x=-5$ on the way towards the barrier. To calculate the final
velocity we used the times it reached $x=-5$ and $x=-11$ for the
reflected soliton and $x=11$ and $x=21$ for the transmitted one. This
way we were calculating the velocities in the region where there were
no effects of the barrier.  In Fig.~2 we plot the modulus of the
outgoing versus incoming velocity.

\begin{figure}
\begin{center}
\includegraphics[width=8cm]{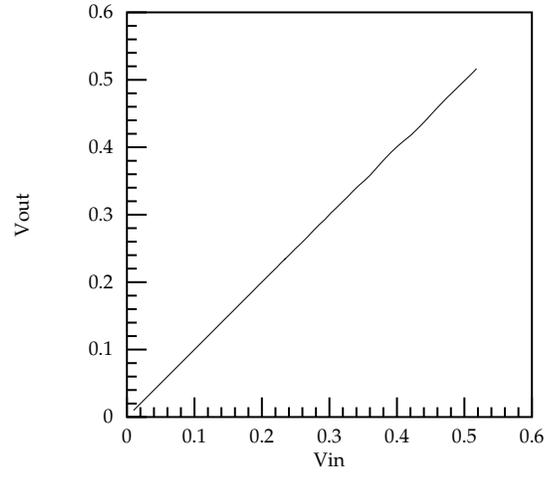}
\caption{ Soliton velocities, $x$ axis - incoming, $y$ axis - outgoing}
\end{center}
\end{figure}
We note that the curve is essentially a straight line suggesting a
simple linear relationship. In fact, the values
of the incoming and outgoing velocities differ from each other in 3 or 4th decimal
points. So we can treat them as equal, within the numerical accuracy of our 
procedure.

The value of the critical velocity can be estimated by observing the 
trajectory of the soliton; if the soliton makes it to the 'top of the barrier',
which is around $x\sim 2.25$, then it goes through, otherwise, it is
reflected. In Fig.~3 we present plots of the time dependence of the soliton
position in two cases; one corresponding to the velocity of just below 
the critical value, namely, $v\sim 0.2426$ and one just above at $v\sim 0.2428$. 

\begin{figure}
\begin{center}
\includegraphics[width=8cm]{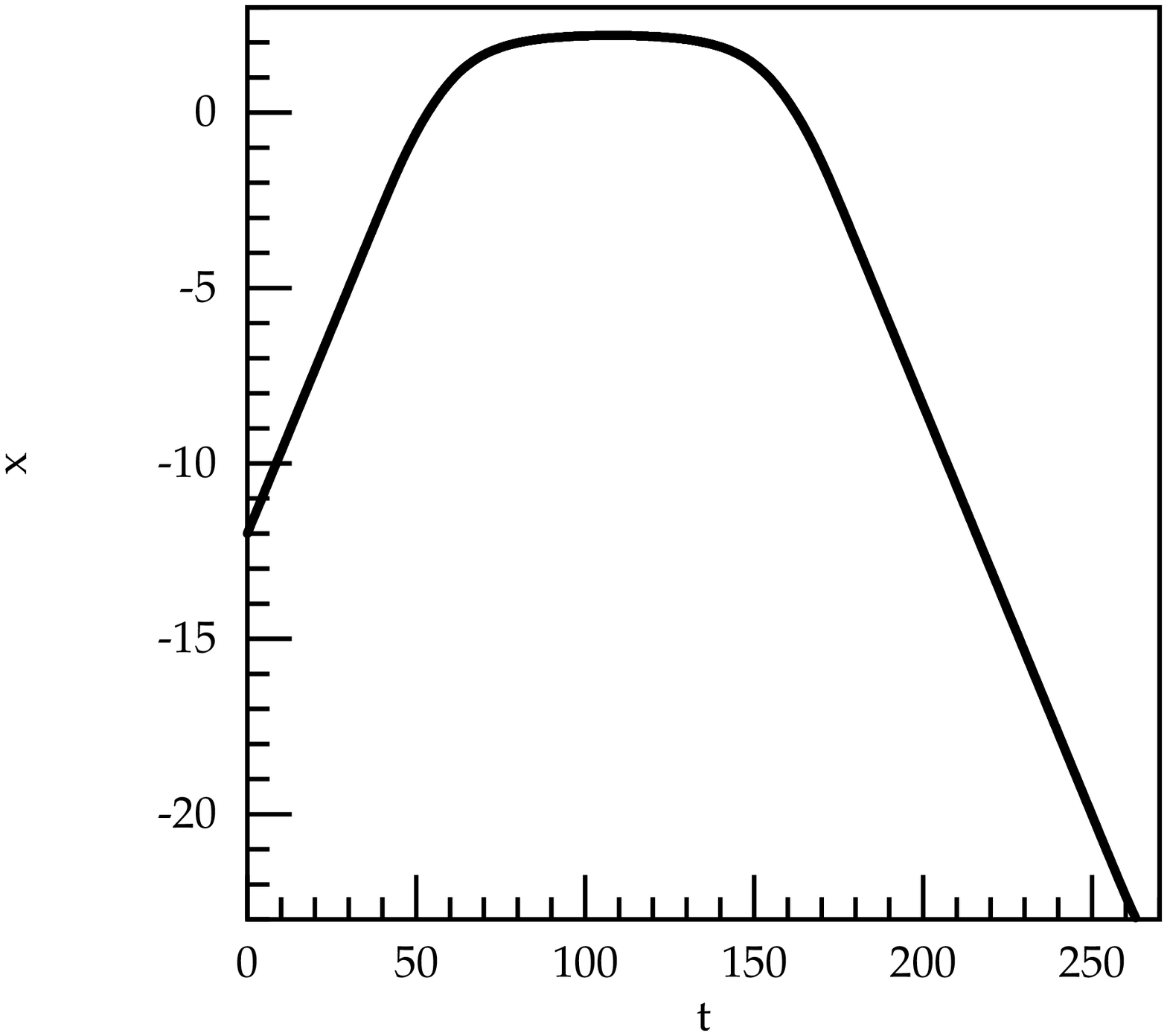}
\includegraphics[width=8cm]{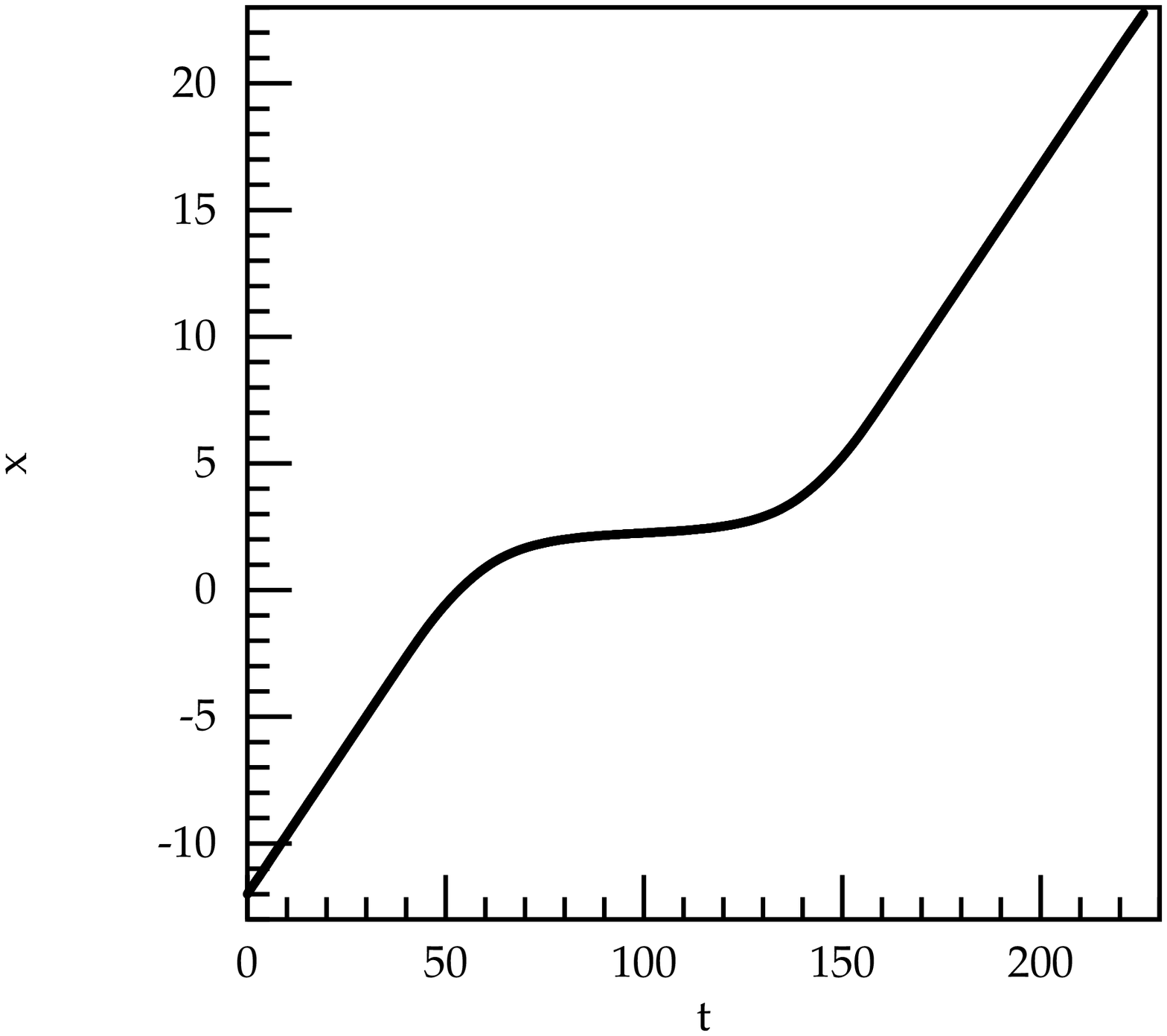}
\caption{ Time dependence of the position of the soliton
initially placed at $x=-12$: left $v<v_{cr}$, right $v>v_{cr}$}
\end{center}
\end{figure}

We note that, in each case, the slopes of the curves before and after
the soliton interacted with the barrier, are the same. 
Thus we see that the scattering is completely elastic; essentially
no energy is lost during the climb of the barrier and because of this
the velocity is unchanged, to the degree of accuracy of our calculation.

Looking at the static configuration with $\mu=0.7$ we note that its energy is 
1.968 which is exactly the energy that the incoming soliton has to have
to be able to transfer all its kinetic energy into the potential 
energy - to be able to climb the barrier and to be  transmitted. 
Thus $v_{cr}$ is determined by the energy that the incoming
soliton corresponding to $\mu=0.6$ has to have so that its
energy corresponds to the energy of the static soliton of $\mu=0.7$.

Note that one can give the following nonrelativistic argument which
supports our claims.  First note that the mass of the soliton is given
by the total energy ($c=1$) so $M\sim 1.912$. Then ${Mv_{cr}^2 \over
2}\sim 1.912*0.05890329/2=0.05631$ which is in agreement with the
difference of energies $\delta E=1.968-1.1912\sim 0.056$.

This assumes an elastic behaviour and this is what we have
seen in our numerical simulations. Thus in the case of the bump - the whole
kinetic energy of the incoming soliton is converted into the potential
energy
of the soliton at the top of the barrier and then the soliton
can be transmitted elastically.

\subsection{Hole}
We placed the hole in the same place as the barrier and we took
$\mu_1-\mu_0=0.5$. Next, as in the barrier case, we placed the initial
soliton at various values of $x$ and calculated its energy. 
This has told us what the hole is like as seen by
the soliton. The calculated energies are shown in Fig.~4 
\begin{figure}
\begin{center}
\includegraphics[width=8cm]{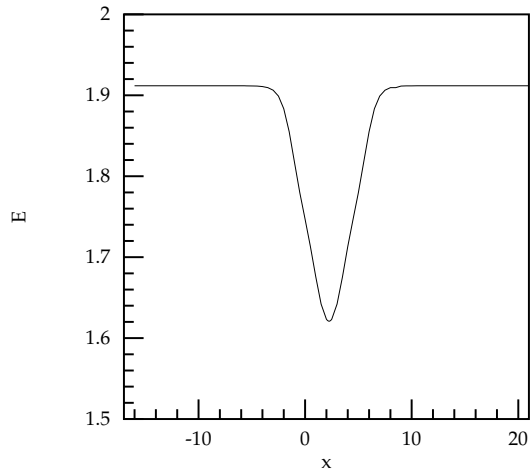}
\caption{ Energies of our basic soliton as a function of its 
position}
\end{center}
\end{figure}
Once again we see that the effective hole as seen by the soliton is quite smooth. We have performed 
many simulations and have found that, again, there exist a critical
velocity $v_{cr}$. Above $v_{cr}$ the soliton is transmitted and below 
it falls in and becomes trapped in the hole.
 
The critical velocity is around $v_{cr}\sim 0.155$.
Solitons started off from $x=-12$ fall into the hole
and stay there oscillating and gradually losing their energy by emitting
radiation. Two typical trajectories 
are shown in Fig.~5. The left picture shows a trapped soliton
with initial velocity $v\sim 0.09 $, ie below $v_{cr}$, and the right one shows a
transmitted soliton with initial velocity $v\sim 0.16$.

\begin{figure}
\begin{center}
\includegraphics[width=8cm]{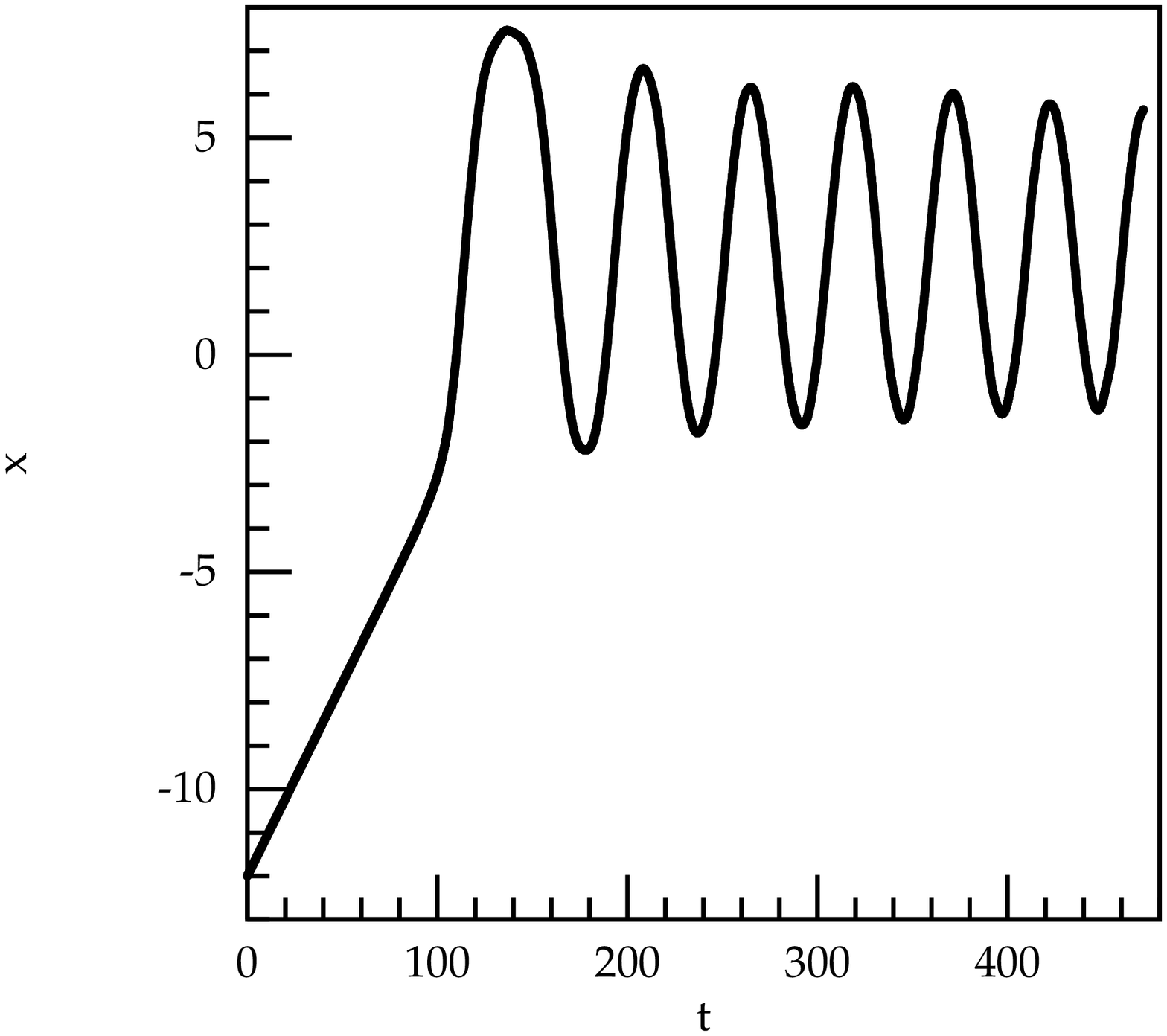}
\includegraphics[width=8cm]{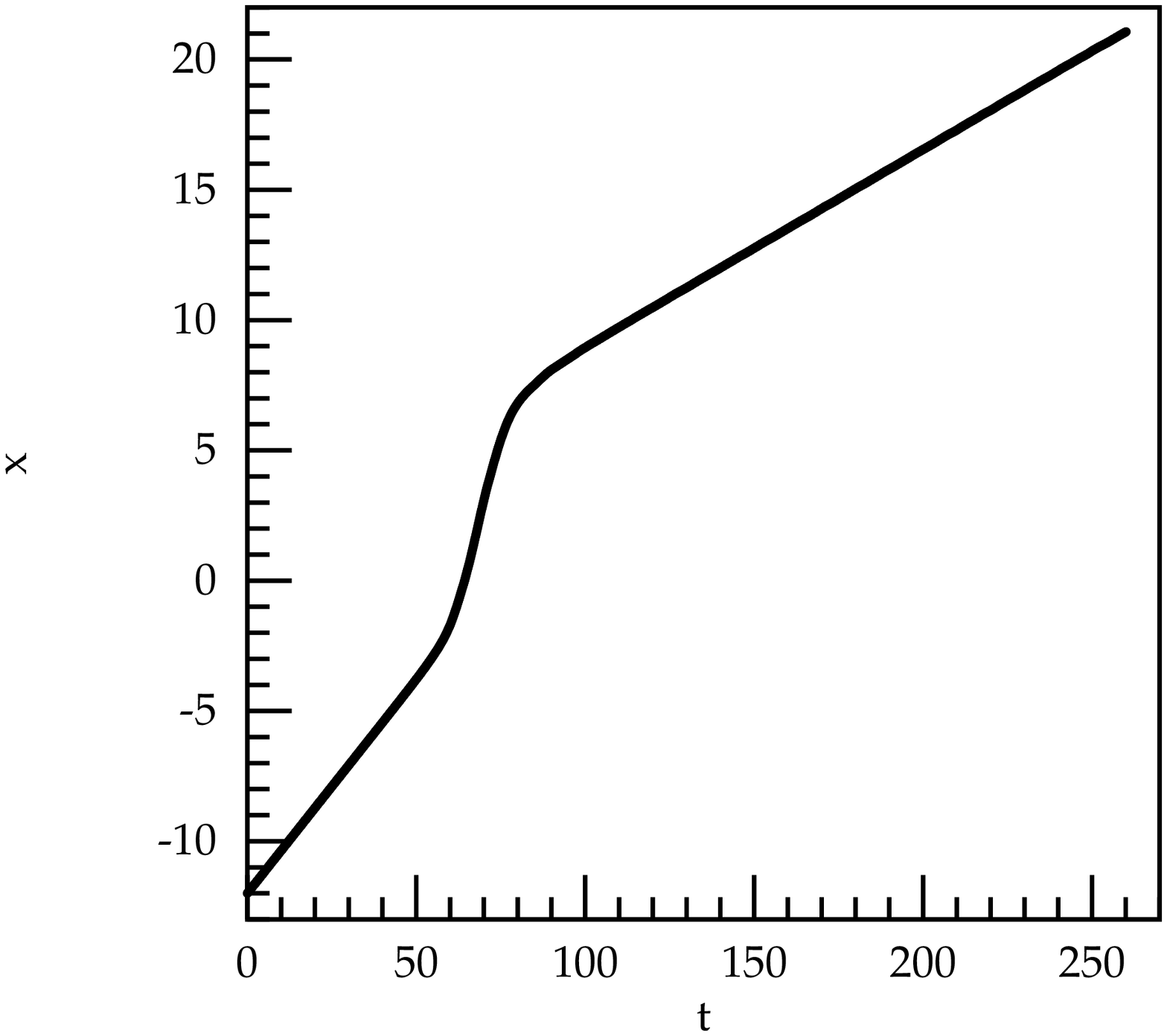}
\caption{Time dependence of the position of the soliton
initially placed at $x=-12$: left $v<v_{cr}$, right $v>v_{cr}$ }
\end{center}
\end{figure}
 
 For velocities above $v_{cr}$ we have a transmission, but this time
 the outgoing soliton is much slower. 
 In Fig.~6 we present the plot of the outgoing velocities as a function
 of the incoming ones.
 
 \begin{figure}
\begin{center}
\includegraphics[width=8cm]{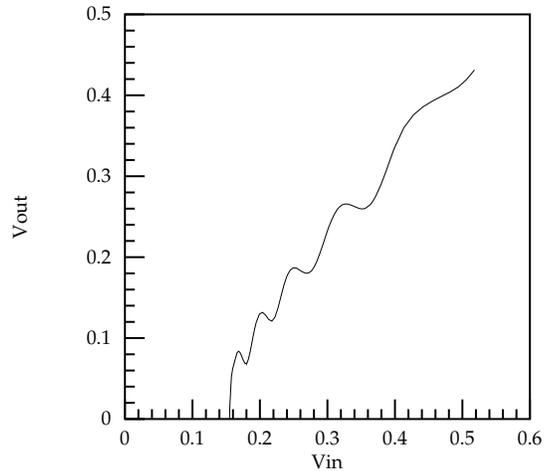}
\caption{Velocities of transmitted solitons, $x$ axis - incoming, $y$ axis - outgoing }
\end{center}
\end{figure}

We note not only the significant decrease of the velocity but also
the oscillations in the values of the outgoing velocities.

We have looked at the details of the scattering but we have not 
succeeded in revealing the origin of the oscillations.
We have rerun some simulations with absorption of the waves at the boundaries; 
the results were essentially the same suggesting 
that the origin of the oscillations has nothing to do with any waves
of emitted radiation bouncing off the boundaries.

Thus - the origin of the effect is associated with the interaction of the soliton with the hole 
and the emission of the radiation  during this process.
To check this we have looked at the time spent in the hole
and its relation to the time needed for the soliton to traverse the hole
had the soliton moved with the initial or the final velocity.

 In Fig.~7a we plot the time spent by the soliton in the hole, as a function of the incoming
 velocity. In Fig.~7b we plot three curves; all as a function of the initial
 velocity. They present the difference of the time the soliton spent in the hole
 from which is subtracted the time needed to traverse the hole with the initial velocity,
 the same with the final velocity and the same with the average velocity.
 
 We see a dramatic difference in all three expressions: 
\begin{figure}
\begin{center}
\includegraphics[width=8cm]{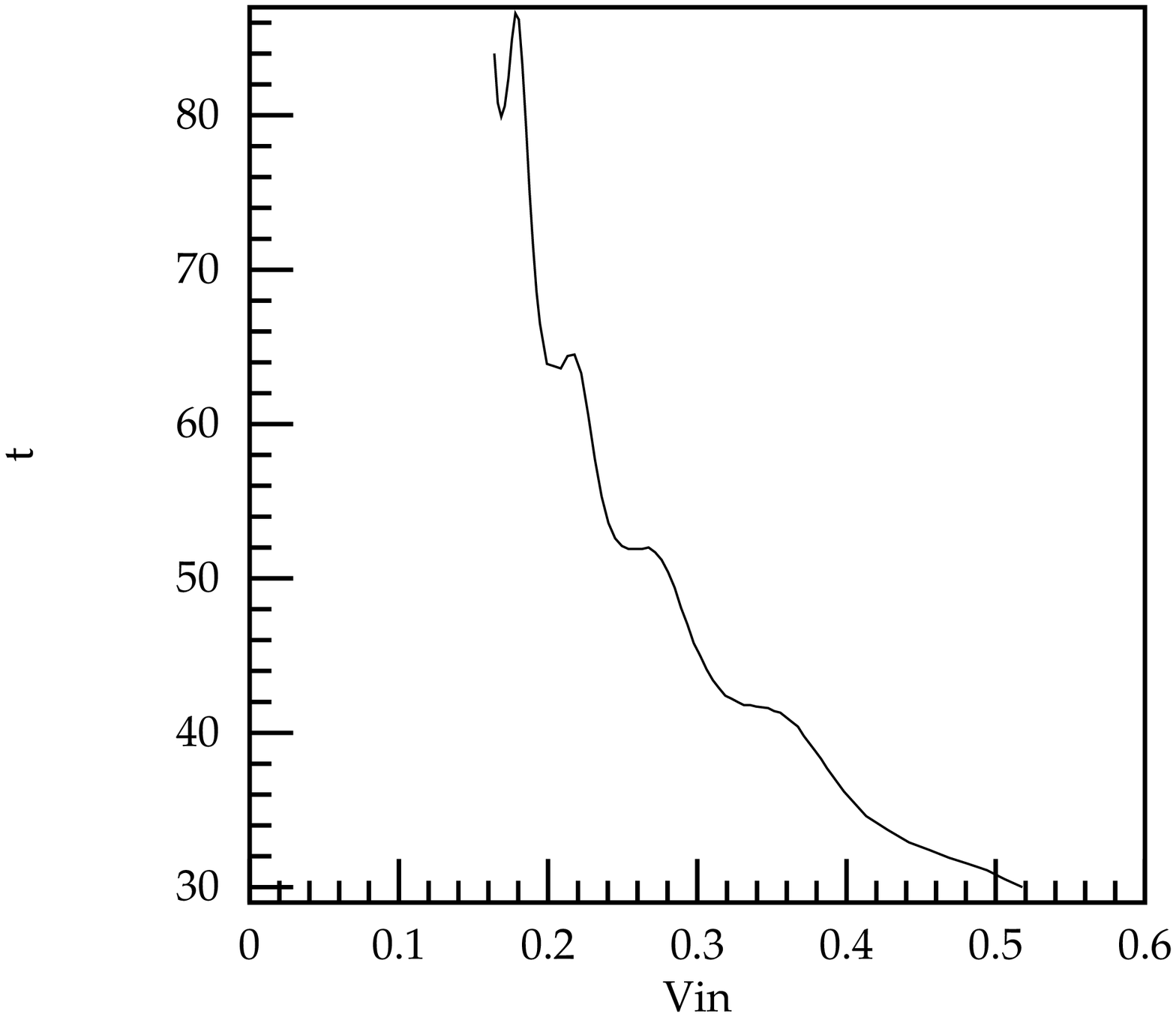}
\includegraphics[width=8cm]{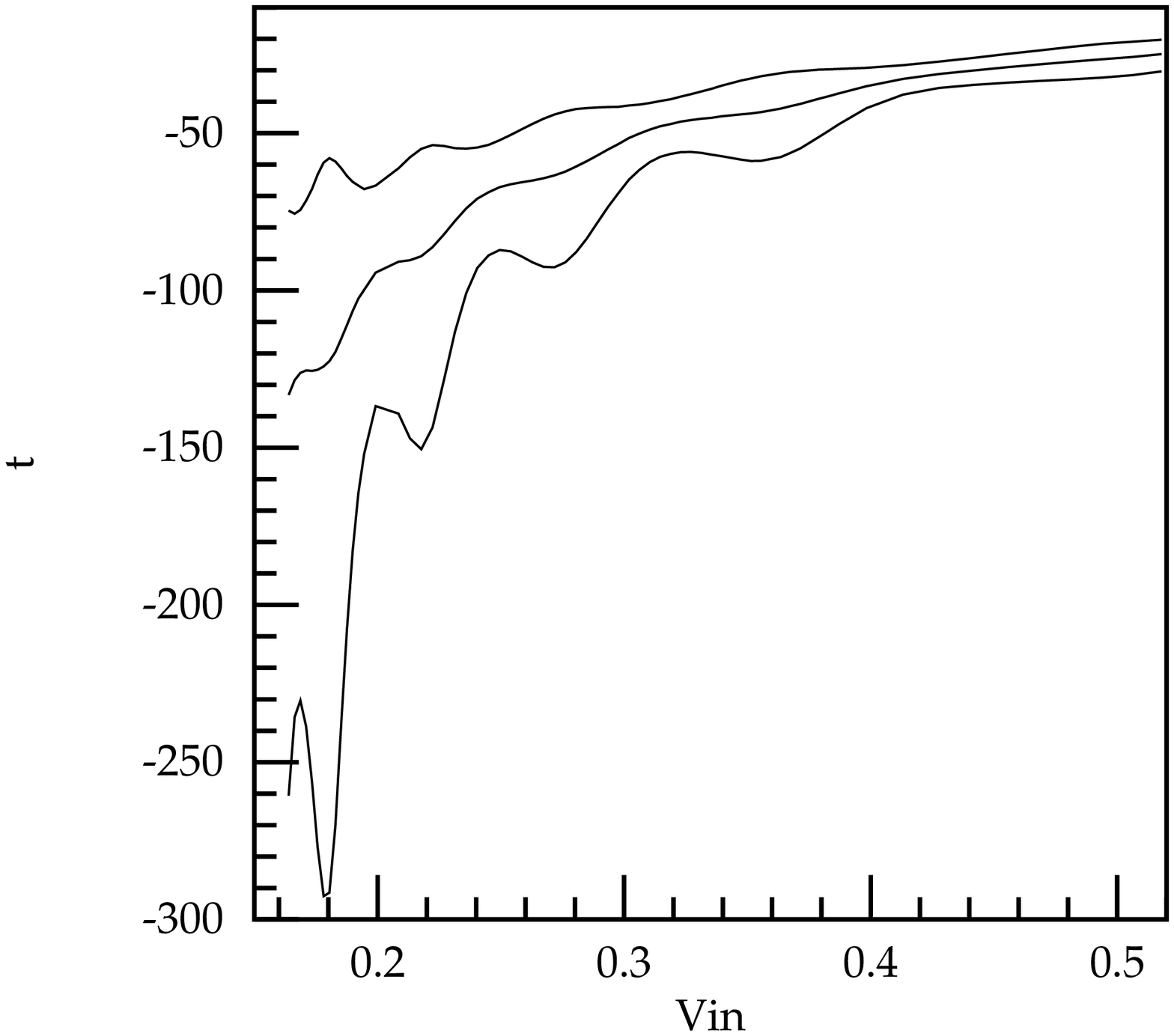}
\caption{left - time spent in the hole; right - time in the hole minus
the time needed to traverse the hole with various velocities. Top curve - initial
velocity, middle average, bottom - outgoing one }
\end{center}
\end{figure}

It is clear that the time in the hole is related to the oscillations of the 
velocities. It is also clear that as the incoming velocity is much larger
than the outgoing velocity the expression for the time spent
in the hole - the time needed to traverse the hole at a given velocity is the 
lowest when we take the largest velocity, i.e., the incoming velocity.
It is also clear that the potential hole increases the velocity - so
all our expressions in Fig.~7b are negative.

Looking at the curves in Fig.~7b we see that the oscillations in
outgoing velocity are pretty much reproduced by the time spent by the
soliton in the hole and almost eliminated when we take the middle
curve in Fig.~7b. This suggests that during the motion through the
hole the soliton velocity changes in a complicated way (gradually
producing the oscillations). If, in the plots in Fig.~7b we use the
initial velocity the oscillations show up in the difference of times;
if we use the final velocity we overcompensate and we have the
opposite oscillations; taking the average velocity accounts for most
of the oscillations leaving a curve which is much smoother. This
curve, however, does show the extra velocity in the hole, which of
course is related to the incoming velocity.

Looking at the curve we note that its features are as expected; we
note a slow rise with with the increase of $v_{inc}$, which is
consistent with the loss of energy in the hole.

Note that this time the process is very different from the situation of a 
barrier. In the barrier case the incoming soliton has some kinetic 
energy which it has to convert to the potential one to rise up the
barrier. This seems to be done elastically, with essentially no
radiation. In the hole case the soliton has a kinetic energy and as it
moves into the hole it gains extra energy as well. This time the extra 
energy is partially converted into the increased speed of motion but some
of it is got rid of via radiation. In fact, this is what we have seen
in our numerical simulations which did show some radiation.
Then, when it reaches the other side of the hole and it starts `climbing out of
the hole' the soliton may not have enough energy to get out. Hence,
for low speeds it gets trapped in the hole.

Of course, we do not really understand why the soliton does not convert all its extra 
energy into the increased kinetic energy and why the velocities
give us the very interesting patterns that can be seen from the plot in Fig.~6.
Clearly there must be an internal mode of the soliton that gets
excited and which radiates the excess energy.

This point requires further study.

\section{Conclusions}

We have looked at a system involving a topological soliton in two dimensions
and a potential, of both a barrier and a hole-type.

When the soliton was sent towards the barrier its behaviour resembles that
of a particle. Thus at low energies the soliton was reflected by the barrier
and at higher energy it was transmitted. The scattering process was very
elastic.
During the scattering the kinetic energy of the soliton was gradually
converted into the energy needed to `climb the barrier'. If the soliton
had enough energy to get to the `top' of the barrier then it was transmitted,
otherwise it slid back regaining its kinetic energy.

Note that the soliton size is related to the parameters of the model
and so depends on $\mu$. Hence, during the climb of the barrier, the soliton
altered its size (it decreased a little) - to fit the local value of $\mu$;
when it got through, or slipped back, its size returned to it original value.
This is what one would expect in an elastic scattering and this is what
we saw in the numerical simulations.
In fact, the soliton size oscillated a little, around its `correct' value
and the amplitude of these oscillations has not changed much
during the scattering process and the final oscillations resembled the original
ones.

In the hole case, the situation was very different. This time, the
soliton gained an extra energy as it entered the hole. Some of this
energy was converted into kinetic energy of the soliton, some was
radiated away.  So when the soliton tried to `get out' of the hole it
had less kinetic energy than at its entry and, when this energy was
too low it remained trapped in the hole.  During the scattering
process, like in the case of a barrier, the soliton size changed and
its oscillations increased significantly. Afterwards they stayed like
this - with much higher amplitude of oscillations than before.  Hence
the increase in oscillations is related to the inelasticity of the
process and the emitted radiation.

The scattering of a topological soliton on a hole is thus reminiscent
of a classical particle under the influence of friction. We have not
found any indication of nonclassical reflection as in the case of
Nonlinear Schr{\"o}dinger solitons \cite{Joachim}. However, in
contrast to a classical particle with simple velocity-dependent
friction, we have found interesting oscillations hinting at some
underlying resonant mechanism.

We have looked in detail at the behaviour of the solitons during their
scattering process and, so far, have not found a satisfactory
explanation of the observed 'oscillations' in the outgoing velocities
(in the case of transmission by the hole).

Clearly, this is related to the properties of the radiation - the only
plausible explanation we can find is that the radiation is sent out in
separate bursts. This is very much what we saw in the actual simulations.
In Fig.~8 we present a plot of the density of the kinetic energy
seen in a typical simulation.

\begin{figure}
\begin{center}
\includegraphics[width=10cm]{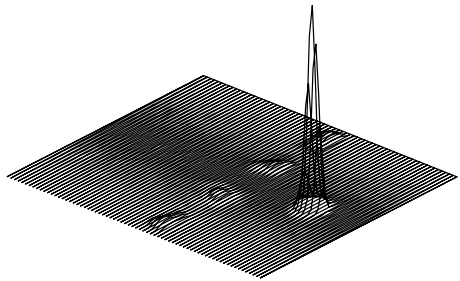}
\caption{The density of kinetic energy in a simulation just after 
the soliton has passed the hole. $T=150$, $v_{in}=1.9165$ }
\end{center}
\end{figure}
 
Looking at the plot we see distinctive waves of energy moving along
the 'trough' ({\it i.e.} in the $y$ directions). The `high' peak is
the kinetic energy density of the moving soliton.
 
  
Clearly we need to find an explanation of the observed radiation 
pattern. But, this lies outside the scope of this paper.

\section*{Acknowledgements}
This investigation was initiated following a discussion between one of the authors (WJZ)
and Sergej Flach. We want to thank Sergej for his interest, support and fruitful
discussions.

WJZ wants to thank the Max Planck Institute in Dresden for its hospitality.

{}

\end{document}